\documentclass[useAMS,usenatbib]{mn2e}
\usepackage{amsmath,amssymb,graphicx,color,longtable}
\newcommand{\mg}{MG 0414$+$0534}
\newcommand{\kmin}{$\kappa_{\rm min}$}
\newcommand{\ksad}{$\kappa_{\rm sad}$}
\newcommand{\gmin}{$\gamma_{\rm min}$}
\newcommand{\gsad}{$\gamma_{\rm sad}$}

\title[Microlensing constraints on lenses and discs]{A joint microlensing analysis of lensing mass and accretion disc models}
\author[G. Vernardos]{
	G. Vernardos$^{1}$\thanks{E-mail: gvernard@astro.rug.nl}
	\\
	$^{1}$Kapteyn Astronomical Institute, University of Groningen, PO Box 800, NL-9700AV Groningen, the Netherlands\\
}
\date{Accepted XXX. Received YYY; in original form ZZZ}
\pubyear{2017}

\begin{document}
\label{firstpage}
\pagerange{\pageref{firstpage}--\pageref{lastpage}}
\maketitle

\begin{abstract}
Microlensing of multiply imaged quasars is a unique probe of quasar structure, down to the size of the accretion disc and the central black hole.
Flux ratios between close pairs of images of lensed quasars can be used to constrain the accretion disc size and temperature profile.
The starting point of any microlensing model is the macromodel of the lens, which provides the convergence and shear values at the location of the multiple images.
Here I present a new approach of microlensing modelling independently of the macromodel of the lens.
The technique is applied to the close pair of images $A_1$ and $A_2$ of \mg, for a set of flux ratios with large variation with respect to wavelength.
The inferred accretion disc size and temperature profile measurements, as well as the smooth matter fraction at the location of the images, are quite robust under a wide range of macromodel variations.
A case of using purely microlensing data (flux ratios) to constrain the macromodel is also presented.
This is a first application of the technique on a fiducial system and set of flux ratios; the method is readily applicable to collections of such objects and can be extended to light curve and/or imaging data.
\end{abstract}


\section{Introduction}
\label{sec:intro}

Cosmological microlensing observations constitute a unique probe of the structure of lensing galaxies and lensed quasars.
Understanding the dark (smooth) and stellar (compact) matter components in galaxy-scale systems is an open issue and has many implications for studying their formation and evolution scenarios \citep[e.g.][]{Conroy2009,Moster2010,Behroozi2010}.
To this end, using strong gravitational lenses has been valuable \citep[e.g.][]{Treu2010,Oguri2014,Leier2016}.

In the case of the lensed source being a quasar, microlensing can be employed to unveil the structure of the accretion disc and the geometry of the emitting regions in the vicinity of the supermassive black hole \citep[e.g.][]{Dai2010,Morgan2010,Guerras2013,ODowd2015}.
This, in turn, can be used to understand the growth of the black hole \citep[e.g.][]{Rosas-Guevara2015,Terrazas2017} and its relation to the quasar host galaxy and its environment via feedback mechanisms \citep[e.g.][]{Bourne2017,Cowley2018}.

For any quasar to be microlensed, it has to be first multiply imaged by a foreground lensing galaxy (the `macrolens', or just `lens').
The positions of the images, any extended lensed features of the background quasar host galaxy, and other available data (e.g. time delays or flux ratios between the images) can be used to construct a mass model for the lens \citep[e.g. see][]{Keeton2001a}.
Such models describe the total mass of the lens, and provide the convergence, $\kappa$, and shear, $\gamma$, fields.
However, the degeneracy between its baryonic and dark matter components remains. 
To lift this degeneracy, the light profile of the lens can be used to measure the smooth matter fraction, $s$ (equation \ref{eq:smf}), as a function of radius \citep{Oguri2014,FoxleyMarrable2018}.
This approach, however, is accompanied by the large uncertainty in the stellar initial mass function, used to convert the light into the mass distribution.
The individual values of $\kappa$, $\gamma$, and $s$, at the locations of the multiple images are the primary parameters for setting the microlensing properties.

Incoming light rays from the background quasar are further deflected by several stellar-mass microlenses existing within the lens and lying along the line of sight to the quasar images.
The presence of such collective deflections creates a network of caustics which can be described by a magnification map \citep{Kayser1986}.
The properties of these maps (e.g. the caustic density, orientation, etc) depend mainly on $\kappa$, $\gamma$, and $s$, which set the mass density of the essential grainy (i.e. stellar in this case) mass component.
The final result is a microlensing-induced time-dependent magnification on the source, uncorrelated between its observed (macro) images.
Analyzing observations using microlensing techniques can provide a measurement for $s$ \citep{Schechter2002}, which can otherwise be only approximated as explained in the previous paragraph.
This has been done using microlensing light curve data \citep[e.g.][]{Chartas2009,Dai2010,MacLeod2015} or microlensing flux ratios \citep[e.g.][]{Bate2011,Pooley2012,Jimenez2015a}.

Besides $\kappa$, $\gamma$, and $s$, the size of the source with respect to the caustics plays an important role: the smaller the background source, the more prominent the microlensing induced brightness variations will be.
It is currently thought that quasar accretion discs are hotter in their innermost regions and cool down further from the central supermassive black hole.
The standard thin-disc model \citep[][]{Shakura1973} predicts a power-law dependence of the temperature as a function of radius, with the power-law index fixed to 3/4.
This is easily transformed into a size-wavelength relation, making discs appear bigger in long (red) and smaller in short (blue) wavelengths.
This wavelength-dependent microlensing effect has been used to constrain quasar accretion discs \citep{Bate2008,Floyd2009,Jimenez2014,Rojas2014,Bate2018}.

All microlensing studies so far have employed the `traditional' two-stage modeling approach.
Firstly, a lens mass model is fitted to the imaging data and the individual values of $\kappa$, $\gamma$ are extracted for each image.
Secondly, a set of microlensing magnification maps is produced as a function of $s$ (or other parameters like the microlens masses, proper motions, etc).
A series of flux ratios or light curves are produced from the maps for different accretion disc profiles and compared to the observations (in the case of light curves, the time delay between the macro-images has to be used to correct the data first).
The very high computational cost associated with generating magnification maps for different parameters \citep{Bate2012}, and the adequately constrained lens mass models from imaging data justify the choice of using fixed values for $\kappa$, $\gamma$.

The possibility of inferring microlensing constraints, and their robustness, on the lens mass model has not been investigated before.
Conversely, studies of the effect of lens model variations/uncertainties on accretion disc constraints, or $s$, inferred by microlensing have been very limited \citep[e.g. see][]{Vernardos2014c}.
The main reason behind this is the computationally demanding task of producing magnification maps for many different combinations of $\kappa$, $\gamma$, and $s$.

The new approach presented in this work assesses the robustness of the derived $s$ and accretion disc constraints with respect to the lens mass model (i.e. the $\kappa$, $\gamma$).
The feasibility of using purely microlensing data and methods in providing constraints to the lens mass model is also examined.
Any constraints on $\kappa$, $\gamma$ coming from the macromodel (i.e. having them as fixed parameters) are therefore dropped, and they are treated as free parameters instead.
Although a computationally more intensive task as a whole, the bulk of the effort, which is computing magnification maps, can be avoided by using the GERLUMPH\footnote{\tt http://gerlumph.swin.edu.au} collection of maps \citep{Vernardos2014a,Vernardos2014b}, whose uniform and extensive coverage of the $\kappa$, $\gamma$, and $s$ parameter space makes it ideal for such an application.
The model and its implementation, as well as the choice of a fiducial system to apply it, are described in Section \ref{sec:method}.
Results are presented in Section \ref{sec:results}, followed by discussion and conclusions in Section \ref{sec:discuss}.

\section{Method}
\label{sec:method}

The geometry of the multiple images of a lensed source is well understood and can be reproduced by relatively simple elliptical mass models.
Understanding the absolute brightness of the individual images is a more complicated task: one has to know the intrinsic brightness of the source, its variability, and the time delays between the images, which are much more sensitive to the exact lensing mass configuration \citep{Kochanek2006}.
Although these effects can be mitigated by using the relative brightness, i.e. the flux ratios of the images, one still has to take into account microlensing and substructure in the lens \citep{Mao1998,Metcalf2001}.
In the absence of such contaminating effects, lensing theory provides a useful result: close image pairs in a fold configuration are expected to have magnifications of roughly the same magnitude \citep{Schneider2006} and therefore an expected magnification ratio of unity (a similar rule holds for a cusp configuration of the images).

The new technique presented in Section \ref{sec:model} is applied to one such system, i.e. the close image pair of \mg, introduced in Section \ref{sec:pair}.
This pair, as expected, consists of a saddle-point ($A_2$) and a minimum ($A_1$) image, which are labeled accordingly in the following.
The specific details of applying the model to the data are presented in Section \ref{sec:implementation}.

\subsection{Model description}
\label{sec:model}
The new approach introduced in this work consists of allowing the $\kappa$, $\gamma$ values for the images to vary.
The relative contribution of the smooth component to the total mass density is assumed to be the same for both images.
This assumption is justified by the close separation of the image pair and its azimuthal orientation around the center of the lens (i.e. the images are found at roughly the same direction and distance from the lens center).
This approximation has been widely used in the literature \citep[e.g.][]{Bate2011,Jimenez2014,Bate2018} as it greatly facilitates the computations, and, to first order, produces meaningful results.

The size of the accretion disc as a function of wavelength is given a parametric power-law form:
\begin{equation}
\label{eq:profile}
r = r_0 \left(\frac{\lambda}{\lambda_0}\right)^{\nu},
\end{equation}
where $r_0$ is the size at the fiducial wavelength $\lambda_0 = 1026$\AA, which together with the power-law index, $\nu$, constitute the two free parameters of the disc model.
The size $r$ is matched to the half-light radius, $r_{1/2}$, of a circularly symmetric (face-on) Gaussian brightness profile for the source\footnote{\citet{Mortonson2005} have shown that the actual shape of such a profile does not play an important role, and it is the size of the half-light radius that matters for the purposes of microlensing.}.
The absolute values of the brightness are unimportant because, as explained above, only flux ratios are examined in this work.

The general form of the Bayesian posterior probability distribution is:
\begin{equation}
\label{eq:bayes}
P(\boldsymbol{p}|\boldsymbol{d},\boldsymbol{\eta}) = \frac{ L(\boldsymbol{d} | \boldsymbol{\eta} , \boldsymbol{p}) Pr(\boldsymbol{p}) }{ E(\boldsymbol{d}|\boldsymbol{\eta}) },
\end{equation}
where $\boldsymbol{p}$ is the vector of the free parameters for this model (\kmin, \gmin, \ksad, \gsad, $s$, $r_0$, $\nu$), and $\boldsymbol{d}$ is the data from Table \ref{tab:ratios}.
$\boldsymbol{\eta}$ is a vector of parameters that we may choose to keep fixed (either $\kappa_{\rm min},\gamma_{\rm min}$ or $\kappa_{\rm sad},\gamma_{\rm sad}$, see Section \ref{sec:results}; other parameters that one may wish to keep track of could be added here, e.g. the average mass of the microlenses, etc) and is omitted in the rest.
$Pr$ is the prior probability of the parameters $\boldsymbol{p}$, and $E$ is the Bayesian evidence.
The likelihood term, $L$, for a fixed set of parameters $\boldsymbol{p}$ is given by:
\begin{equation}
\label{eq:like}
L (\boldsymbol{d}|\boldsymbol{p}) = \sum_{\rm k=1}^{N} L_{\rm k} = \sum_{\rm k=1}^{N} {\rm exp} \left( - \frac{\chi^2_{\rm k}}{2} \right),
\end{equation}
as the sum over all the chi-squared realizations:
\begin{equation}
\label{eq:chi2}
\chi^2_{\rm k} = \sum_{\rm i=1}^{4} \left( \frac{f^{\rm obs}_{\rm i} - f^{\rm sim}_{\rm i,k}}{\sigma_{\rm i}} \right)^2,
\end{equation}
where the index $i$ corresponds to the observed flux ratios, $f^{\rm obs}$, and their uncertainties, $\sigma$, as a function of wavelength, and the index k corresponds to our simulated flux ratios, $f^{\rm sim}$.
Obtaining $f^{\rm sim}$, the strategy of finding $L$ as a function of the free parameters $\boldsymbol{p}$, and the priors used are described in the next sections.

\subsection{The close pair of \mg}
\label{sec:pair}
The new approach presented here is applied to images $A_1$ (minimum) and $A_2$ (saddle-point) of the quadruply imaged quasar \mg~which have a separation of $\delta\theta \approx 0.4$ arcsec \citep{Hewitt1992}.
Due to a deviation (anomaly) from the expected magnification ratio of unity in the UV and optical, which persists in infrared and radio observations (where any microlensing effect is expected to be negligible), this particular system has been the focus of several studies of possible substructure in the lens \citep{Mao1998,Dalal2002,Minezaki2009,MacLeod2013}.

Additionally, a number of microlensing analyses have been performed on this system: \citet{Bate2008,Bate2011} and \citet{Blackburne2011} find a temperature profile of the quasar accretion disc which is consistent with the thin disc model, while \citet{Pooley2007} find a size larger than expected.
Recently, \citet{Bate2018} have used Hubble Space Telescope (HST) data to measure an accretion disc with size $\mathrm{ln}(r_{\rm 0}) < 1.07$ ($r_{\rm 0}$ in light days) and slope $\nu=2.1^{+0.6}_{-0.6}$ (modelled after equation \ref{eq:profile}), marginally larger than thin disc theory expectations.

In this study, we adopt the microlensing flux ratio data obtained by \citet{Bate2018}, shown in Table \ref{tab:ratios}.
We also use the macromodel of \citet[][table 3]{MacLeod2013}, which consists of three components: the main lens, modelled as a Singular Isothermal Ellipsoid (SIE) with external shear, a known companion galaxy, modelled as a Singular Isothermal Sphere (SIS), and an unknown (dark) substructure, also modelled as a SIS.
Based on this macromodel, \citet{Bate2018} computed the values of the convergence and the shear of each image in the pair, hereafter referred to as $\kappa_{\rm ML13},\gamma_{\rm ML13}$ (see Table \ref{tab:means_sdevs}).

\begin{table}
	\centering
	\caption{Flux ratios between images $A_2$ and $A_1$ of {\mg} as a function of observed wavelength, $\lambda$, adopted from \citet{Bate2018}.}
	\label{tab:ratios}
	\begin{tabular}{cc}
		\hline
		$\lambda$ (\AA) & $A_2/A_1$ \\
		\hline
		7612  & 0.34 $\pm$ 0.03 \\
		8436  & 0.42 $\pm$ 0.02  \\
		12486 & 0.66 $\pm$ 0.01 \\
		15369 & 0.76 $\pm$ 0.01 \\
		\hline
	\end{tabular}
\end{table}

Based on the data of \citet{Bate2018}, the macromodel of \citet{MacLeod2013}, and general properties of close image pairs, the following remarks/simplifications can be made.
Firstly, the time delay between the images is expected to be very short \citep[e.g. see][for an analysis of 10 systems, including \mg]{Pooley2007}, and so the quasar can be essentially considered in the same state for both images at the time of observation.
Secondly, the flux in each filter that is coming from regions (and physical scales) beyond the accretion disc (and thus effected differently by microlensing) is minimal; this has been achieved by carefully selecting which HST filters to observe with \citep[see fig. 1 of][]{Bate2018}.
Thus, in the following, the wavelength dependence of the flux ratios is attributed solely to the structure of the quasar accretion disc and its ongoing microlensing.
Lastly, as explained above, in the case of an unperturbed lens mass model and without any differential extinction, the expected magnification ratio would be equal to unity.
However, the presence of substructure in the lens \citep[][]{MacLeod2013} and/or the possible effect of differential extinction \citep[which is harder to correct for as it requires spectroscopic data, e.g.][]{Jimenez2014,ODowd2015} are causing deviations from unity.
These effects are taken into account by setting a baseline magnification ratio of $f_{\rm base} = 0.93\pm0.03$, assumed to be unaffected by microlensing.
This was obtained from the infrared observations and subsequent models of \citet{Minezaki2009} and \citet{MacLeod2013}.

\begin{figure*}
	\includegraphics[width=\textwidth]{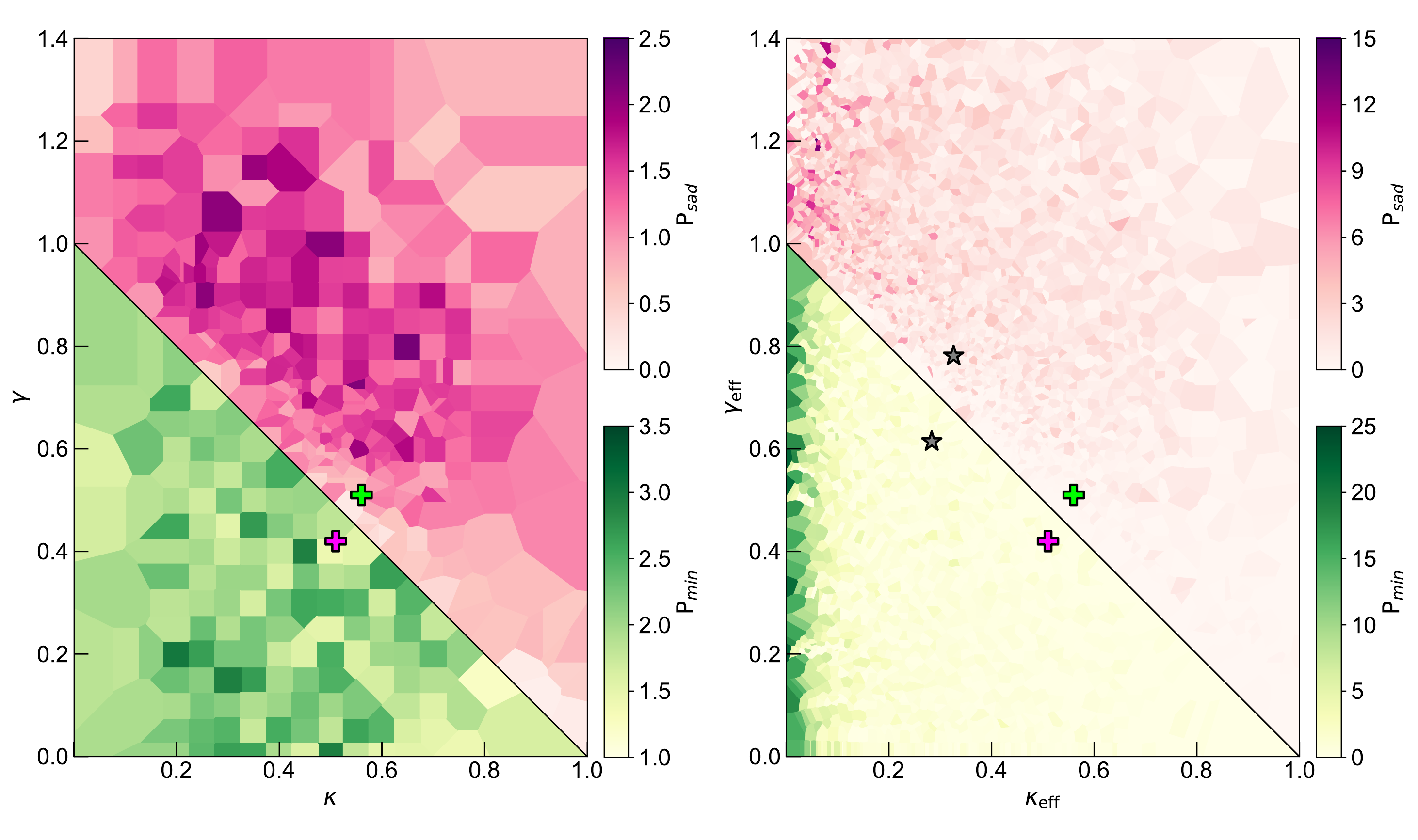}
	\caption{Results for CON6. Left: $\kappa-\gamma$ joint probability density marginalized over $s$, $r_0$, and $\nu$, plotted as shaded Voronoi cells, with a darker (lighter) color indicating a higher (lower) probability. Right: probability density for $\kappa,\gamma,s$, marginalized over the accretion disc parameters $r_0$ and $\nu$, shown in the effective parameter space (using the transformation of equation \ref{eq:eff}). The parameter space in each panel is divided by the critical line (black solid line, see also equation \ref{eq:mag}) separating the saddle-point and minimum regions, above and below it respectively. The likelihood surface shown for the saddle-point is computed while keeping the minimum image fixed to its $\kappa_{\rm ML13},\gamma_{\rm ML13}$ values (indicated by a cross) and vice-versa. The locations of the effective $\kappa_{\rm ML13},\gamma_{\rm ML13}$ are also marked (grey stars), using the value of $s=0.61$ from Table \ref{tab:means_sdevs} (for REF).}
	\label{fig:min_sad_pspace}
\end{figure*}

\begin{figure*}
	\includegraphics[width=\textwidth]{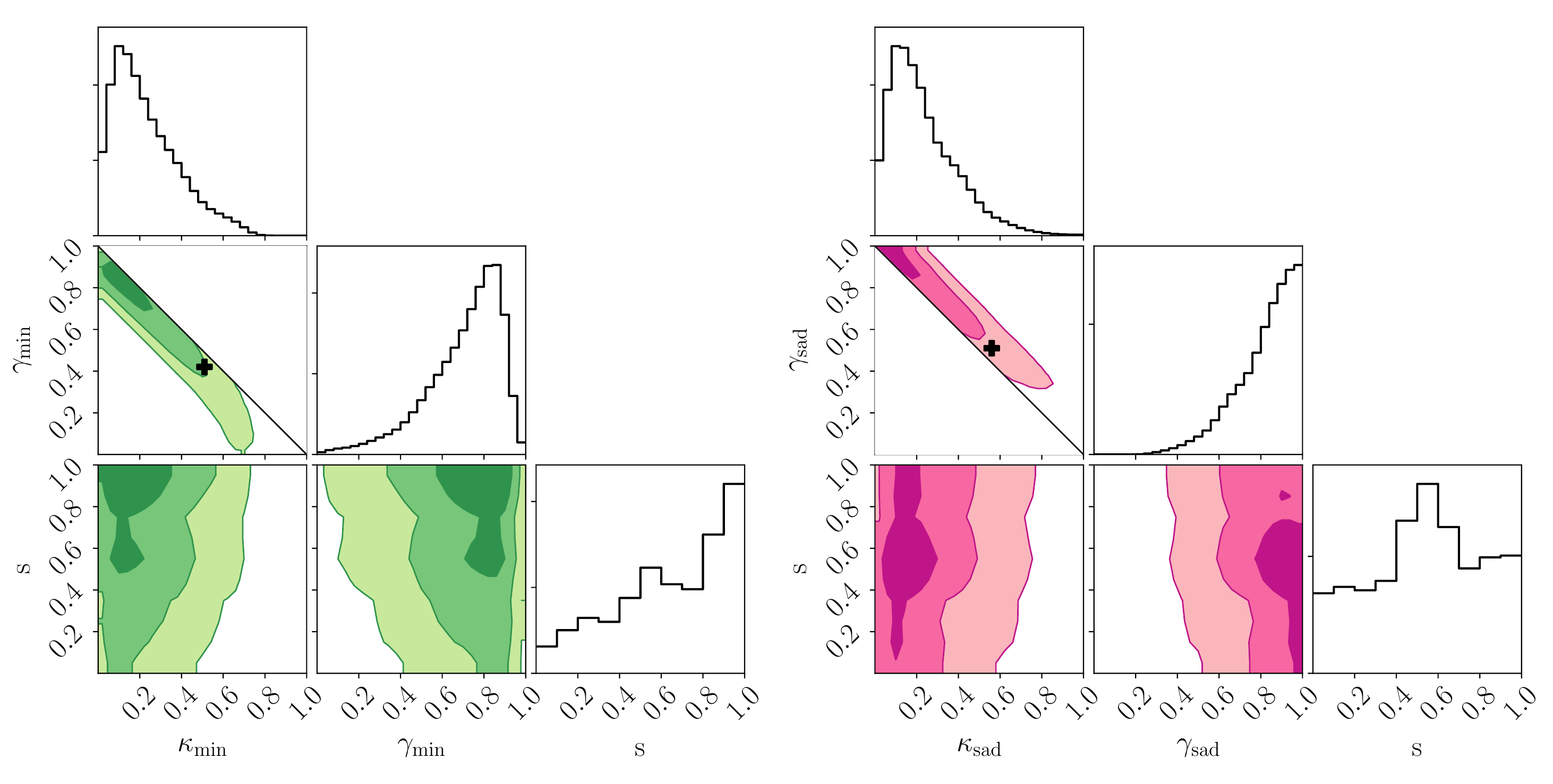}
	\caption{Constrained probability densities and histograms for \kmin, \gmin, $s$ and \ksad, \gsad, $s$ (CON7), marginalized over the accretion disc parameters $r_0$ and $\nu$. Contours are drawn at the 68, 95, and 99 per cent confidence intervals. Crosses indicate the location of the $\kappa_{\rm ML13},\gamma_{\rm ML13}$ values.}
	\label{fig:min_sad_corner}
\end{figure*}

\subsection{Implementation}
\label{sec:implementation}
The macromodel, or lens, parameters consist of the convergences and shears, \kmin, \gmin, \ksad, \gsad, and the smooth matter fraction, $s$, assumed to be the same for both images.
The accretion disc parameters are the size, $r_0$, and power-law slope, $\nu$.
Thus, the model can have up to a total of 7 free parameters.
For the macromodel parameters we adopt the ranges $0<\kappa<1$, $0<\gamma<1.4$, and $0\leq s \leq 0.9$ (the last one in steps of 0.1).
Magnification maps were retrieved from the full GERLUMPH\footnote{Both GD1 and GD3 datasets were used, which are computed on a regular but sparse and an irregular but dense $\kappa,\gamma$ grid respectively. All the maps are available online at: \tt http://gerlumph.swin.edu.au} dataset \citep[see fig. 4 of][and related text for details]{Vernardos2014b}.
The Einstein radius on the source plane, $R_{\rm Ein}$, is set to $3.74 \times 10^{16}$ cm for microlenses with a fixed mass of $1 M_{\odot}$, using $z_S = 2.64$ \citep{Lawrence1995} and $z_L = 0.96$ \citep{Tonry1999} for the measured redshifts of the source and the lens, and a Universe with $H_0 = 70$ km s$^{-1}$ Mpc$^{-1}$, $\Omega_m = 0.3$, and $\Omega_{\Lambda} = 0.7$.
For the accretion disc parameters a regular grid is selected such that ${\rm ln}(r_0) = 0.3 \times j$ for $j=0\dots11$ and $\nu = 0.25 \times i$ for $i=0\dots15$, following \citet{Jimenez2014}.

The remaining procedure is almost identical to the one presented in \citet{Bate2018}.
For each combination of ln($r_0$) and $\nu$, a set of two-dimensional, symmetric, face-on, Gaussian profiles \citep[see][]{Mortonson2005} are generated for the accretion disc in each wavelength of Table \ref{tab:ratios}.
The half-light radius of each profile is $r_{1/2} = 1.18 r$, where $r$ comes from equation (\ref{eq:profile}), i.e. it is the standard deviation of the Gaussian.
The profiles are truncated at $2\times r_{1/2}$, having a total width of $4 \times r_{1/2}$.
Whenever a profile has a total width larger than 16 $R_{\rm Ein}$\footnote{This limit is debatable as the caustics can still have a structure on this scale, depending on the values of $\kappa$, $\gamma$, and thus there could be still some microlensing effect present. See also the discussion in Section \ref{sec:discuss}.} it is regarded as being too large to be affected by microlensing and the flux ratio is assumed to have the baseline value $f_{\rm base}$.

For the rest of the profiles, in order to get the simulated flux ratios, $f^{\rm sim}$, to be used in equation (\ref{eq:chi2}), a convolution with each magnification map has to be carried out first.
Due to the convolution edge effects, instead of the entire convolved maps only a central `effective' part of them is used.
The size of this effective map is determined by the largest profile, i.e. the one in the reddest wavelength $\lambda = 15369$\AA, e.g. for ${\rm ln}(r_0),\nu = (0.3,1)$ equation (\ref{eq:profile}) gives $r = 5.23 \times 10^{16}{\rm cm} \approx 1.4 R_{\rm Ein}$ and the effective map size is $18.4 R_{\rm Ein}$ (from the 25 $R_{\rm Ein}$ GERLUMPH maps).
Magnification values are drawn from a square grid of $10^4$ points in each effective map, producing $10^8$ simulated flux ratios in each wavelength.
Hence, first the $\chi^2$ term of equation (\ref{eq:chi2}) is calculated, and then the sum of equation (\ref{eq:like}), that has $N=10^8$ terms, is computed.

The analysis and results presented below are based on relative posterior probabilities, therefore, the computationally demanding calculation of the evidence term in equation (\ref{eq:bayes}) is disregarded.
Such a computation would be meaningful in the case of comparisons between different physically motivated models for the lens or the disc, which is feasible within the general formulation introduced above, but out of the scope of this paper.
Because of this, the terms likelihood and probability are used interchangeably in the following.
Fixed grids are adopted for the exploration of the parameter space, leaving the use of other, more elaborate and efficient sampling techniques, such as Markov Chains, Gibbs sampling, or other optimizers, for future work.

Finally, all the priors were chosen to be flat, except for $r_{\rm 0}$ that has a logarithmic prior \citep{Bate2018}.
One could argue that $s$ should have a logarithmic prior as well, since it is a multiplicative parameter:
\begin{equation}
\label{eq:smf}
\kappa_{\rm *} = (1-s)\kappa, 
\end{equation}
where $\kappa_{\rm *}$ is the convergence in compact matter.
In the next section the results were computed using both priors for $s$.

\section{Results}
\label{sec:results}

\begin{figure}
	\includegraphics[width=0.5\textwidth]{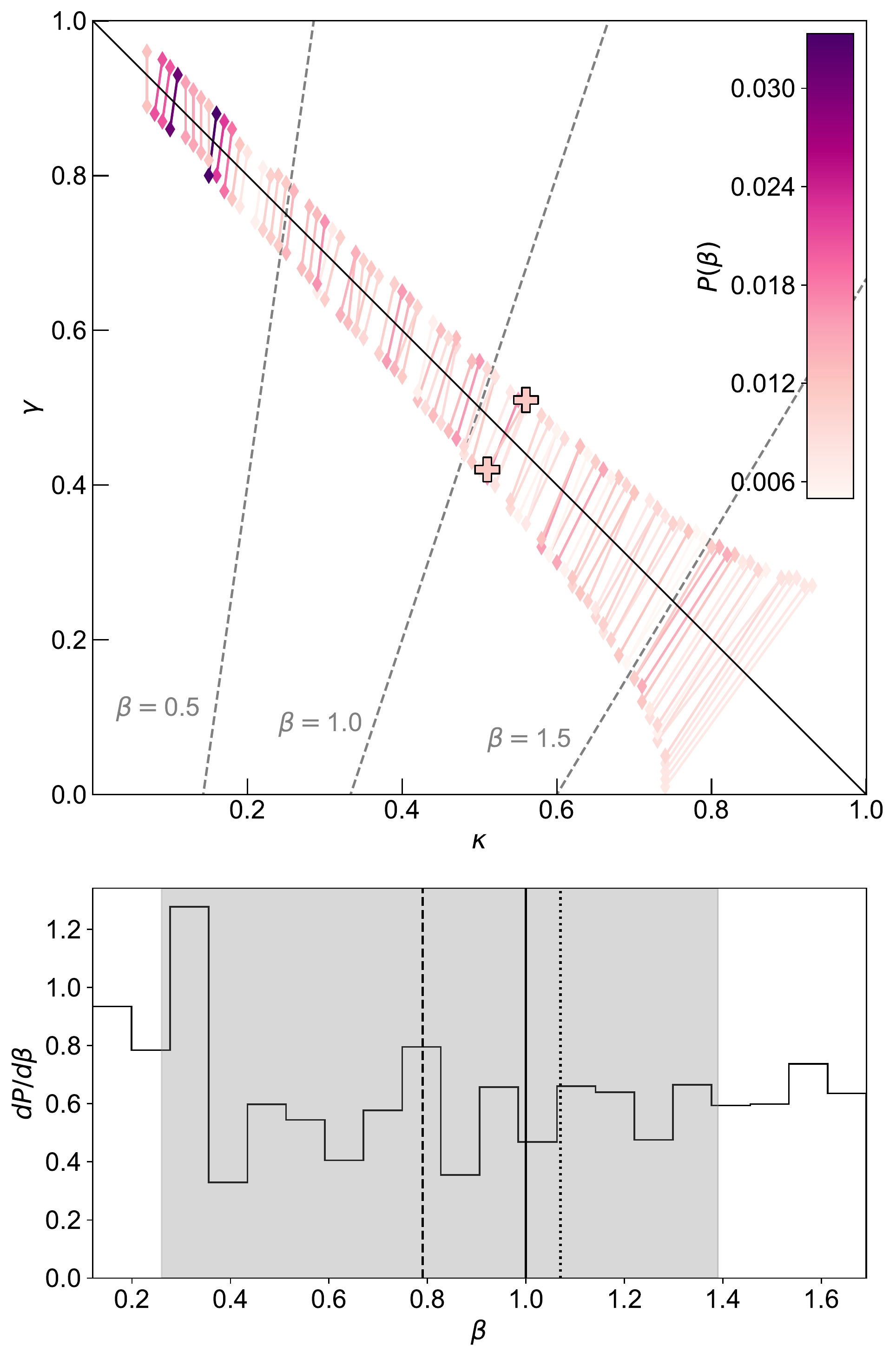}
	\caption{Top: pairs of minimum - saddle-point images in the parameter space, with darker (lighter) colors indicating higher (lower) probability (CON8). Crosses mark the $\kappa_{\rm ML13},\gamma_{\rm ML13}$ values (REF). The dashed lines correspond to equation (\ref{eq:slope}) for specific values of $\beta$, as in fig. 1 of \citet{Witt1995}. Bottom: probability density of the slope of a fiducial spherical potential for the lens, obtained by fitting equation (\ref{eq:slope}) to the pairs shown in the top panel. The vertical solid line indicates the case of an isothermal potential ($\beta = 1$), the dotted line shows the slope value obtained by fitting $\kappa_{\rm ML13},\gamma_{\rm ML13}$ for the two images, the dashed line the value of $\beta = 0.79$ obtained from CON8 and the grey shaded area its 68 per cent confidence interval.}
	\label{fig:pairs}
\end{figure}

\begin{figure}
	\includegraphics[width=0.5\textwidth]{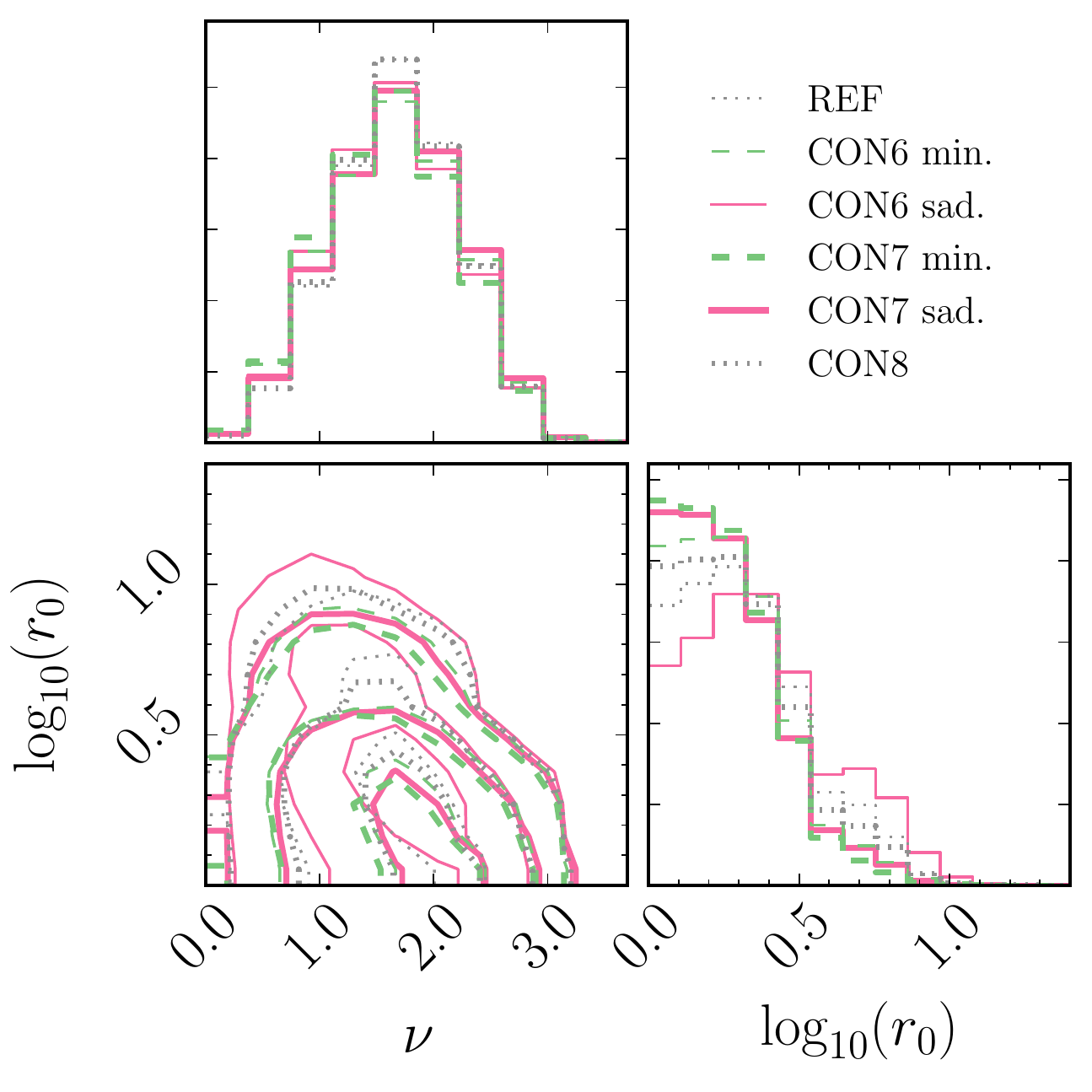}
	\caption{Probability density and histograms for the accretion disc parameters $r_0$ (in light-days) and $\nu$, corresponding to the size of the accretion disc at the rest wavelength $\lambda_0 = 1026$\AA and its power law dependence on wavelength (see equation \ref{eq:slope}). All the different set of results presented here are shown, marginalized over $\kappa,\gamma$ and $s$ whenever applicable. Contours are drawn at the 68, 95, and 99 per cent confidence intervals.}
	\label{fig:disc_corner}
\end{figure}

The model presented in the previous section has a total of 7 free parameters (\kmin, \gmin, \ksad, \gsad, $s$, $r_0$, $\nu$).
A completely unconstrained variation of the $\kappa,\gamma$ values for both images, together with the rest of the parameters, is a computationally demanding task, especially when using fixed grids to explore the parameter space.
The results presented in this section are divided into sets having different constraints.
The flux ratio data, shown in Table \ref{tab:ratios}, are used in all cases and provide 4 constraints to the model.
The $\kappa,\gamma$ of each, or both, of the images, are allowed to vary freely or under some constraint:
\begin{itemize}
	\item{REF:} this is a benchmark, or reference, set, keeping both $\kappa_{\rm min},\gamma_{\rm min}$ and $\kappa_{\rm sad},\gamma_{\rm sad}$ fixed to the corresponding $\kappa_{\rm ML13},\gamma_{\rm ML13}$ values (the number of constraints is 8). The same parameter values and setup is used as in \citet{Bate2018}.
	\item{CON6} allowing either $\kappa_{\rm min},\gamma_{\rm min}$ or $\kappa_{\rm sad},\gamma_{\rm sad}$ to vary freely in the parameter space while keeping the other fixed to the $\kappa_{\rm ML13},\gamma_{\rm ML13}$ values (the number of constraints is 6).
	\item{CON7:} same as in the previous set, but in this case the varying $\kappa,\gamma$ are constrained by equation (\ref{eq:mag}) in order to reproduce the magnification given by the $\kappa_{\rm ML13},\gamma_{\rm ML13}$ values (the number of constraints is 7).
	\item{CON8:} varying both $\kappa_{\rm min},\gamma_{\rm min}$ and $\kappa_{\rm sad},\gamma_{\rm sad}$ under the constraint of reproducing the magnification given by the $\kappa_{\rm ML13},\gamma_{\rm ML13}$ values (using equation \ref{eq:mag}) and matching to a given slope of a fiducial spherical potential for the lens (using equation \ref{eq:slope}; the number of constraints is 8).
\end{itemize}

The magnification is obtained from the lens equation \citep[e.g. see][]{Schneider2006} as:
\begin{equation}
\label{eq:mag}
\mu = \frac{1}{(1-\kappa)^2 - \gamma^2}.
\end{equation}
This equation is used to define the critical line, i.e. the locus of points in the $\kappa,\gamma$ plane, corresponding to a straight line, where the magnification goes to infinity.
The critical line serves also as a division between the minimum ($\mu > 0$) and saddle-point ($\mu < 0$) regions of the parameter space (see Fig. \ref{fig:min_sad_pspace}).

In the left panel of Fig. \ref{fig:min_sad_pspace}, we show the probability surface from equation (\ref{eq:bayes}) as a function of \kmin,\gmin~and \ksad,\gsad~respectively, marginalized over the remaining parameters $s$, $r_0$, and $\nu$ (CON6).
This is equivalent to the likelihood surface of equation \ref{eq:like} under the use of flat priors and examining relative probability values.
A total of 140 (300) combinations of $\kappa_{\rm min},\gamma_{\rm min}$ ($\kappa_{\rm sad},\gamma_{\rm sad}$) is shown, selected randomly in the parameter space.
The resulting grid is irregular and a generic `pixel' nees to be associated with each probed location.
Here, the $\kappa,\gamma$ plane is partinioed in Voronoi cells, which enclose the points closer to a specific probed location than to any other location.
Another choice of partitioning could be the Delaunay triangulation, however, this would associate 3 grid points rather than 1, as is the case with a Voronoi cell that is closer to the notion of a pixel centered on a measurement.
The Voronoi and Delaunay tesselations are the dual of each other, and are unique.
For further marginalization over either $\kappa$ or $\gamma$ (e.g. to obtain the expectation values and confidence intervals shown in Table \ref{tab:means_sdevs}), the likelihood is weighted by the area of each Voronoi cell.

The transformation provided by \citet{Paczynski1986}:
\begin{equation}
\label{eq:eff}
\kappa_{\rm eff} = \frac{(1-s)\kappa}{1-s\kappa} \, , \, \gamma_{\rm eff} = \frac{\gamma}{1-s\kappa},
\end{equation}
is a consequence of the mass-sheet degeneracy \citep{Falco1985}, and reduces the three macromodel parameters $\kappa,\gamma,s$ to only two: the effective convergence, $\kappa_{\rm eff}$, and shear, $\gamma_{\rm eff}$, where $\kappa_{\rm eff}$ is now due only to compact microlenses.
This transformation allows the collapsed likelihood\footnote{The likelihood in the left panel of Fig. \ref{fig:min_sad_pspace} is collapsed with respect to $s$; for each $\kappa,\gamma$ it is the sum of all the individual likelihoods for different $s$. Having the likelihood as a function of $\kappa,\gamma$, and $s$ allows to plot the right panel of Fig. \ref{fig:min_sad_pspace}.}, shown in the left panel of Fig. \ref{fig:min_sad_pspace} as a function of $\kappa,\gamma$, to be shown as a function of $\kappa_{\rm eff},\gamma_{\rm eff}$ in the right panel of the same figure.
The transformation introduces a weighting of the probability density by the determinant of its Jacobian matrix:
\begin{equation}
\label{eq:det_eff}
|\mathrm{det} \frac{\partial (\kappa,\gamma)}{\partial (\kappa_{\rm eff},\gamma_{\rm eff})}| = \frac{(1-s\kappa)^3}{1-s}, \quad (\kappa \leq 1).
\end{equation}
For any fixed $\kappa,\gamma$ and a varying $s$, the resulting $\kappa_{\rm eff},\gamma_{\rm eff}$ from equation \ref{eq:eff} lie on straight lines radiating from $(1,0)$.
This is the reason for the higher concentration of points to the left and top of the right panel of Fig. \ref{fig:min_sad_pspace} \citep[see also fig. 1 of][]{Vernardos2014c}.

\begin{table*}
	\centering
	\renewcommand{\arraystretch}{1.5}
	\caption{Expectation values for the macromodel ($\kappa,\gamma,s$) and accretion disc (${\rm ln}(r_0),\nu$) parameters at the 68 per cent confidence interval for the four sets of results introduced in Section \ref{sec:results}. The size parameter $r_{\rm 0}$ is measured in light days. The $\kappa,\gamma$ values for REF are based on \citet{MacLeod2013}. Values that are shown without uncertainties are kept fixed.}
	\label{tab:means_sdevs}
	\begin{tabular}{cccccccc}
		\hline	
		& REF	(both fixed)		& \multicolumn{2}{c}{CON6 (one free)}					& \multicolumn{2}{c}{CON7 (one constrained)} 			& CON8 (both varying)	\\ 
		&							& minimum					& saddle-point				& minimum					& saddle-point 				& 						\\
		\hline
		$\kappa_{\rm min}$	& $0.51^{}_{}$				& $0.41^{+0.17}_{-0.21}$  	& $0.51^{}_{}$				& $0.18^{+0.19}_{-0.11}$ 	& $0.51^{}_{}$				& $0.42^{+0.26}_{-0.30}$	\\
		$\gamma_{\rm min}$	& $0.42^{}_{}$				& $0.23^{+0.27}_{-0.17}$  	& $0.42^{}_{}$				& $0.74^{+0.12}_{-0.23}$ 	& $0.42^{}_{}$				& $0.83^{+0.05}_{-0.30}$	\\
		$\kappa_{\rm sad}$	& $0.56^{}_{}$				& $0.56^{}_{}$				& $0.41^{+0.36}_{-0.29}$	& $0.56^{}_{}$				& $0.17^{+0.21}_{-0.12}$ 	& $0.50^{+0.34}_{-0.36}$	\\
		$\gamma_{\rm sad}$	& $0.51^{}_{}$				& $0.51^{}_{}$				& $1.09^{+0.14}_{-0.27}$	& $0.51^{}_{}$				& $0.86^{+0.10}_{-0.18}$ 	& $0.90^{+0.07}_{-0.28}$	\\
		$s$					& $0.61^{+0.21}_{-0.24}$	& $0.65^{+0.18}_{-0.38}$  	& $0.47^{+0.28}_{-0.31}$	& $0.66^{+0.18}_{-0.38}$ 	& $0.53^{+0.25}_{-0.28}$ 	& $0.62^{+0.20}_{-0.34}$ 	\\
		\hline
		${\rm ln}(r_0)$		& $\leq 1.00$ 				& $\leq 0.86$  				& $0.85^{+0.59}_{-0.53}$ 	& $\leq 0.79$ 				& $\leq 0.82$ 				& $\leq 0.92$  				\\
		$\nu$				& $1.65^{+0.50}_{-0.52}$	& $1.63^{+0.53}_{-0.60}$  	& $1.59^{+0.54}_{-0.52}$ 	& $1.57^{+0.55}_{-0.56}$ 	& $1.65^{+0.52}_{-0.57}$ 	& $1.64^{+0.50}_{-0.52}$ 	\\
	\end{tabular}
\end{table*}

Next, a constrained rather than free variation of the $\kappa,\gamma$ values for each image is examined (CON7).
Two new parameters are used, namely, the magnification (equation \ref{eq:mag}) and the displacement along a constant magnification contour, $t_\mu$.
In this way, the observationally motivated constraint on $\mu$ is easily achieved by allowing its value for one image to vary slightly with respect to the fixed magnification of the other image.
Varying $\mu$ between 0.9 and 0.96 $\times f_{\rm base}$ (the baseline magnification without microlensing) in steps of 0.01, and $t$ on fixed intervals, creates a rectangular regular grid for both parameters.
Transforming between $(t,\mu)$ and $(\kappa,\gamma)$ is trivial, however, there is a volume, or weight, associated with each resulting $\kappa,\gamma$ location due to the coordinate transformation:
\begin{equation}
\label{eq:det}
|\mathrm{det} \frac{\partial (\kappa,\gamma)}{\partial (\mu,t)} |= \frac{[(1-\kappa)^2 + \gamma^2]^2}{2\gamma}.
\end{equation}
To obtain the probabilities on this new grid of $\kappa,\gamma$ under the assumed constraint on $\mu$, the likelihood values of CON6 (left panel of Fig. \ref{fig:min_sad_pspace}) are interpolated using the natural neighbour interpolation technique \citep{Sibson1981}.
In Fig. \ref{fig:min_sad_corner}, the constrained probability distributions of $\kappa,\gamma,s$ are shown (CON7), multiplied by the correct weight and marginalized over the accretion disc parameters $r_0$ and $\nu$, for varying \kmin,\gmin (left panel) and \ksad,\gsad (right panel).
As expected, the $\kappa-\gamma$ joint probability contours follow the shape of constant magnification contours [e.g. see Fig. \ref{fig:pairs}, fig. 1 of \citet{Witt1995}, or fig. 7b of \citet{Vernardos2014c}].

In the top panel of Fig. \ref{fig:pairs} we show 100 pairs of images, colored according to their probability (CON8).
In this case, the assumption of keeping one of the two images fixed to $\kappa_{\rm ML13},\gamma_{\rm ML13}$ has been dropped, but the constraint of the pair having a magnification ratio of $f_{\rm base}$ has been retained.
\citet{Witt1995} have investigated singular spherical potentials for the lens having a convergence as a function of radius $r$ of the form:
\begin{equation}
\label{eq:potential}
\kappa(r) = \frac{\beta}{2} \bigg( \frac{b}{r} \bigg)^{2-\beta},
\end{equation}
where $\beta$ is the slope of the mass distribution (or potential, with $\beta = 1$ for a SIS model), and $b$ is a scaling factor related to the Einstein radius of the lens, and derived the theoretical result:
\begin{equation}
\label{eq:slope}
\gamma = \frac{4 - \beta}{\beta} \kappa - 1.
\end{equation}
We have used this relation to loosely correlate each pair shown in Fig. \ref{fig:pairs} with the slope of such a fiducial potential for the lens (a practical reason for this `looseness' is the finite and irregular grid of available magnification maps in the $\kappa,\gamma$ parameter space).
By fitting equation (\ref{eq:slope}) to each pair, the probability density of the slope $\beta$ is derived and shown in the bottom panel of Fig. \ref{fig:pairs}.
The expectation value of $\beta$ is $0.79^{+0.60}_{-0.53}$ at the 68 per cent confidence interval (which is not really meaningful, given the flatness of the distribution in the lower panel of Fig. \ref{fig:pairs}).
Transforming this to probability distributions for $\kappa$ and $\gamma$ (as in the last column of Table \ref{tab:means_sdevs}), the following weights have to be used:
\begin{align}
\label{eq:beta_conversion}
|\frac{\partial \kappa}{\partial \beta}| = \frac{(\kappa+\gamma+1)^2}{4(\gamma+1)}, \\
|\frac{\partial \gamma}{\partial \beta}| = \frac{(\kappa+\gamma+1)^2}{4\kappa}.
\end{align}

The marginalized probabilities of the accretion disc parameters $r_0$ and $\nu$ are shown in Fig. \ref{fig:disc_corner} for all sets of results.
Interestingly, the shape of the probability contours and histograms is almost identical.

Finally, the expectation values for the accretion disc and the lens parameters from all four sets of results are shown in Table \ref{tab:means_sdevs}.
Introducing a logarithmic prior on $s$ (as discussed in Section \ref{sec:implementation}) has a minor effect on these values: slightly lower values are preferred for the derived \kmin~and \ksad, values between 0.3 and 0.4 are preferred for $s$, and slightly higher values between 1.7 and 1.8 are preferred for $\nu$.
However, in both cases the derived values are consistent within their confidence intervals.

\subsection{The computations}
The most computationally demanding part of the simulations undertaken in this paper is generating the microlensing magnification maps for a wide range of $\kappa$, $\gamma$, and $s$.
However, this task has been already accomplished by the GERLUMPH parameter survey, which has made available more than 70,000 magnification maps in the targeted part of parameter space \citep[see][]{Vernardos2014a,Vernardos2014b}.
The number of individual magnification maps used in the case of a fixed image was 10: a single $\kappa,\gamma$ location with 10 different values of $s$.
A set of 1400 maps (140 \ksad,\gsad~locations) were used for a varying minimum image, and 3000 maps (300 \kmin,\gmin~locations) for a varying saddle-point.
To obtain the probability of the 100 pairs shown in Fig. \ref{fig:pairs}, 2000 maps were used.
The total number of magnification maps used is 6400, which would have taken approx. 1,830 days to generate on a single Graphics Processing Unit (GPU), or just 29 days using the GPU-Supercomputer for Theoretical Astrophysics Research (gSTAR).
For comparison, the remaining part of the computations, i.e. the convolutions between maps and source profiles described below, took 10 days on gSTAR.

All the results share a common grid of the accretion disc parameters $r_0$ and $\nu$.
This grid contains 192 unique combinations, which, from equation (\ref{eq:profile}), produce 768 different accretion disc sizes\footnote{The possible case of a combination of $r_0$, $\nu$, and $\lambda$ resulting in practically the same $r$ from equation (\ref{eq:profile}) is disregarded.}.
From these, only the 209 sizes that correspond to accretion disc profiles smaller than $16\times R_{\rm Ein}$ - the adopted no-microlensing limit - were convolved with magnification maps to extract simulated flux ratios, while the rest have been given a fixed ratio equal to $f_{\rm base}$.

A total of 26,752,000 convolutions between $10,000^2$-pixel maps and profiles were performed (maps for both images had to be convolved with the same profile), using multiple GPUs on gSTAR over a period of 10 days.
Our final results consist of 12,288,000 likelihood evaluations (equation \ref{eq:like}), for each of which we computed $10^8$ $\chi^2$ terms either by calculating $f^{\rm sim}$ in equation (\ref{eq:chi2}) as described, or by setting it equal to $f_{\rm base}$.

\section{Discussion and conclusions}
\label{sec:discuss}

Despite the extreme variations in $\kappa$, $\gamma$, leading to dramatically different magnification maps with respect to caustic structure and magnification probability distribution, in all the examined cases the same accretion disc constraints are derived, as shown in the last two rows of Table \ref{tab:means_sdevs} and in Fig. \ref{fig:disc_corner}.
This apparent independence of the accretion disc on the macromodel supports the findings of \citet{Bate2018}: the derived accretion disc properties appear to be tightly connected to the observed data, in this case, the large chromatic variations of the flux ratios.
The macromodel seems to be playing an insignificant role, at least for \mg~examined here and the given extreme chromatic variation of the flux ratios \citep{Bate2018}.

The accretion disc constraints of Table \ref{tab:means_sdevs} are consistent with \citet{Bate2008} for the size and the slope parameters of equation (\ref{eq:profile}), while for the slope the agreement with \citet{Bate2018} is marginal.
The main reason for this is that they used maps with a width of $100 R_{\rm Ein}$, much wider than the $25 R_{\rm Ein}$ maps used here, allowing for the inclusion of larger sources ($>16 R_{\rm Ein}$) in calculating the likelihood surface of Fig. \ref{fig:disc_corner}.
This and a number of other effects have been identified to influence the derived accretion disc constraints to a smaller or larger extent: the size of the effective map, the value of the baseline ratio, $f_{\rm base}$, and its uncertainty, the number of simulated ratios between maps, and the way these were selected (from pixels on a fixed grid, in random locations, etc).
These potential sources of bias will be examined in future work.

More than half of the matter at the location of the examined image pair is found to be in the form of a smooth component, regardless of the macromodel.
This is not surprising because the multiple images form at the outskirts of the lensing galaxy, where the stellar density is expected to be low.
In fact, higher smooth matter fractions can be invoked to explain the observed flux ratio anomaly, usually manifesting itself as a demagnified saddle-point \citep{Schechter2002,Vernardos2014a}.
The value of $s$ from \citet{Bate2018} is $0.5^{+0.3}_{-0.3}$ (N. Bate, private communication), consistent with the values of Table \ref{tab:means_sdevs}.
\citet{Bate2011} find a value of 0.8 for \mg, \citet{Pooley2012} find a higher value of 0.93, while \citep{Jimenez2015a} find a value of 0.8 by examining a collection of 27 image pairs of lensed quasars.
However, the uncertainty on $s$ (Table \ref{tab:means_sdevs}) is quite large in all cases, indicating basically flat distributions.

Based purely on the microlensing observations, without using any other kind of data, is there anything to be said about the lens mass model?
The inferred values of \kmin,\gmin, and \ksad,\gsad, more often disagree with the macromodel of \citet{MacLeod2013} than agree.
Of course, one has to take into account the largely underconstrained nature of the problem: the model has 7 free parameters and the result sets CON6, CON7, and CON8 use 6, 7, and 8 constraints respectively.
Therefore, the values and confidence intervals derived for $\kappa,\gamma$ in Table \ref{tab:means_sdevs} should be taken cautiously.
In general, for the observed flux ratios in Table \ref{tab:ratios}, and without any information on the macromodel (derived from imaging data), it seems that steeper mass distributions than isothermal are favoured, leading to lower $\kappa$ and higher $\gamma$ values at the location of the close pair of images (see Figs. \ref{fig:min_sad_corner} and \ref{fig:pairs}).

It is interesting to investigate the convergence of the solutions of the model as more observational constraints are used.
The method introduced in this paper would be straightforward to apply by adding more terms in equation (\ref{eq:chi2}) and assuming the flux ratios from different observational epochs are uncorrelated\footnote{This means that the source will have to move across the sky by a distance corresponding to at least its own size. \citet{Mosquera2011b} calculate a median source crossing timescale of 7.3 months based on a sample of 87 lensed quasars.}.
Additionally, the effectiveness of using flux ratios with different (smaller) chromatic variations should be tested.
In fact, if each close pair image configuration can be associated with distinct flux ratio properties, then the solutions should converge to the correct $\kappa,\gamma$.
This will be investigated in future work using mock data for several systems with different $\kappa,\gamma$ \citep[similarly to what is suggested in][]{Bate2018}.

A similar ansatz, i.e. finding the macromodel parameters based on microlensing observables, can be suggested and tested in the case of light curves.
The method presented here can be modified accordingly to use light curve data, and the model expanded to include additional parameters such as the velocities of the observer, source, and lens, etc.
However, this would require a careful selection of priors on the new parameters and an understanding of their effect in the interpretation of the results.
This is another path of exploration spurring from this work.

Finally, it is relatively straightforward to combine the analysis presented here with techniques that fit the macromodel to imaging data; it would be a simple addition of flux ratio and image position $\chi^2$ terms.
Such an approach would be meaningful if the solutions of the method presented here are indeed shown to converge to useful values of $\kappa,\gamma$, and could be proven valuable in disentangling microlensing effects from the presence of substructure in the lens.
Combining this method with imaging data would be easier than with light curves.

In this paper, a joint analysis of the lens macromodel and the accretion disc was performed for the first time, driven solely by microlensing flux ratio data.
The derived accretion disc constraints were proven to be quite robust under broad variations of the $\kappa,\gamma$ for each image.
With the method and machinery presented in this study, one can envisage simultaneous analysis of different kinds of available observations, deriving constraints on the lens mass and accretion disc models of a lensed quasar.
The cornerstone for such multi-component modelling approaches is a readily available collection of magnification maps, like GERLUMPH, which removes the need of the huge amount of computations associated with generating them.
The future for lensing studies driven by a variety of available observational data modelled in the same framework looks promising.

\section*{Acknowledgements}
The author would like to thank C. J. Fluke, L. V. E. Koopmans, and N. F. Bate for providing comments and suggestions on early versions of this work, which improved the final result.
The author is supported through an NWO-VICI grant (project number 639.043.308).
This work was performed on the gSTAR national facility at Swinburne University of Technology.
gSTAR is funded by Swinburne and the Australian Government's Education Investment Fund.

\bibliographystyle{mnras}
\bibliography{kg_constraints}

\end{document}